\documentclass[sigconf]{acmart}

\usepackage{tikz}
\usepackage[scaled]{beramono}
\usepackage{listings}
\newcommand{\bcirc}[1]{%
  \tikz[baseline=(C.base)]\node[draw,circle,fill=black,text=white,
  inner sep=1.1pt] (C) {\footnotesize #1};}
  
\AtBeginDocument{%
  }

\acmConference[ABB Corporate Research Center]{}{Mannheim}{Germany}
\begin{document}

\title[Spec2Control]{Spec2Control: Automating PLC/DCS Control-Logic Engineering\\from Natural Language Requirements with LLMs\\ - A Multi-Plant Evaluation}

\author{Heiko Koziolek, Thilo Braun\\Virendra Ashiwal, Sofia Linsbauer}
\email{firstname.lastname@de.abb.com}
\affiliation{%
  \institution{ABB Corporate Research Center, Germany}
  \city{Mannheim}
  \country{Germany}
}

\author{Marthe Ahlgreen Hansen\\Karoline Gr\o tterud}
\email{firstname.lastname@no.abb.com}
\affiliation{
  \institution{ABB Process Automation Energy Industries Division}
  \city{Oslo}
  \country{Norway}
}

\renewcommand{\shortauthors}{Koziolek et al.}

\begin{abstract}
Distributed control systems (DCS) manage the automation for many industrial production processes (e.g., power plants, chemical refineries, steel mills). Programming the software for such systems remains a largely manual and tedious process, incurring costs of millions of dollars for extensive facilities. Large language models (LLMs) have been found helpful in generating DCS control logic, resulting in commercial copilot tools. Today, these tools are focused on textual notations, they provide limited automation, and have not been tested on large datasets with realistic test cases. We introduce Spec2Control, a highly automated LLM workflow to generate graphical control logic directly from natural language user requirements. Experiments using an open dataset with 10 control narratives and 65 complex test cases demonstrate that Spec2Control can successfully identify control strategies, can generate 98.6\% of correct control strategy connections autonomously, and can save between 94-96\% of human labor. Spec2Control is being integrated into commercial ABB engineering tools, but is also available as an open-source variant for independent validation.
\end{abstract}




\maketitle

\section{Introduction}
A DCS is a computerized system that automates industrial equipment used in manufacturing processes~\cite{Hollender2010}. ABB has been a leader in the 22 BUSD DCS market for 24 years, having installed more than 35.000 systems~\cite{ABB2025}. A large DCS may contain dozens of real-time controllers, hundreds of servers, and control hundreds of thousands of sensors and actuators~\cite{ABB2017}. The brain of such installations is the software, which reads sensor values and calculates actuator setpoints based on control logic written in domain-specific IEC 61131-3 programming languages~\cite{Tiegelkamp2010,ABB2019}.

Programming such systems is referred to as control-logic engineering and incurs high costs due to the induced manual labor~\cite{Gutermuth2014}. A mid-sized production plant may require more than a year of control engineering from a team of 10 engineers~\cite{Hollender2010}. Automation engineers interpret prose text customer requirements and implement them in specialized programming environments~\cite{Hollender2010,Tiegelkamp2010,ABB2019}. Despite measures for reusing functionality across projects, this process remains tedious and error-prone and is one of the most significant cost factors for automation projects~\cite{Gutermuth2014,Hollender2010}.

Recently, automation vendors and researchers demonstrated the capability of modern large language models (LLMs) to generate IEC 61131-3 control logic from natural language specifications~\cite{Koziolek2023,Fakih2024,Liu2024}. This capability could automate parts of the control-logic engineering process. For example, the Siemens Industrial Copilot~\cite{Siemens2025a}, TwinCAT CoAgent~\cite{Beckhoff2025a}, or B\&R Automation Studio Copilot~\cite{BRAutomation2024} can create and explain IEC 61131-3 control logic using chat interfaces. These tools are focused on textual notations and have not been systematically tested for large DCS projects so far.

The contribution of this paper is Spec2Control, the first end-to-end LLM workflow that generates graphical IEC 61131-3 function block diagrams (FBD) directly from customer-written specifications. Spec2Control automates the control logic engineering process beyond interactive chat sessions of previous engineering copilots by autonomously mapping customer specifications to control logic. It includes a dedicated context generation approach, a streamlined pseudo-code LLM notation for graphical FBDs, and performs a custom hierarchical auto-layout. 

While we designed Spec2Control for commercial ABB engineering tools, we have extended its output generation to support open-source tooling (OpenPLC~\cite{Alves2014}), enabling independent validation and enhancement. Based on numerous commercial control narratives, we have constructed an open dataset comprising 10 diverse control narratives, which include 65 test cases for control logic. For these test cases, Spec2Control achieved a flawless control strategy detection using GPT-5, 98.6\% correct strategy connections, and 97.1\% correct alarm mappings, potentially saving 94-96\% time compared to human control engineering.

This paper is structured as follows: Section 2 provides background and explains control-logic engineering in more detail. Section 3 describes Spec2Control as a 7-step workflow, supported by a running example. Section 4 presents our multi-plant testing approach with quantitative and qualitative results. Section 5 highlights threats to validity, followed by Section 6, which discusses related work. Section 7 concludes the paper and sketches future work.

\section{Background}
Control logic engineering refers to the process of programming a real-time automation controller to process sensor values and compute output values for actuators and human operators. While some controllers are directly programmed in C, most control engineers use domain-specific programming languages, notably IEC 61131-3 and IEC 61499. The former standardizes three textual notations: structured text (ST, see Fig.~\ref{fig:st}), ladder diagrams (LD), and instruction lists (IL), and two graphical notations: function block diagrams (FBD, see Fig.~\ref{fig:fbd-example}) and sequential function charts (SFC). Control engineers use these notations, for example, to encode PID (proportional/integral/derivative) control algorithms, interlocks between devices, and alarm procedures. They often employ reusable function blocks with standard algorithms from elaborate and well-tested programming libraries, which they parametrize and connect but do not program from scratch.

\begin{figure}[!htbp]
\footnotesize
\begin{lstlisting}
IF I_TI_101_2 >= 70 THEN (* temperature > 70 degrees *)
  P_101_ON_VAR := FALSE; (* stop pump *)
  E_104_ON_VAR := FALSE; (* stop heat exchanger *)
  CMD_TMR(IN := 2.0);	 (* wait for 2 seconds *)
END_IF;
\end{lstlisting}
\normalsize
\caption{Example code snippet for IEC 61131-3 Structured Text (ST) encoding an interlock in a Pascal-like syntax.}
\label{fig:st}
\end{figure}

\begin{figure}[!htbp]
\center
  \includegraphics[width=\columnwidth]{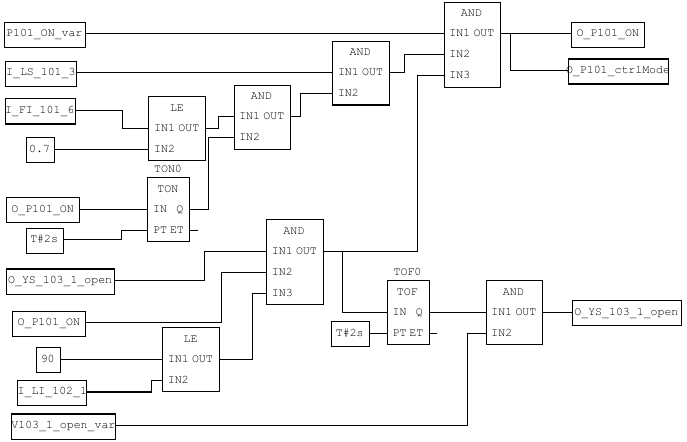}
  \caption{Example for an IEC 61131-3 Function Block Diagram (FBD) encoding simple boolean logic.} 
\label{fig:fbd-example}
\end{figure}

Once implemented and tested, such control logic can comprise thousands of diagrams and hundreds of thousands of function block instantiations~\cite{Krause2007}. The logic gets compiled and deployed on a programmable logic controller (PLC) or Industrial PC~\cite{Hollender2010}. The runtime environment executes control logic cyclically (e.g., every 100-1000 ms). There are also open-source programming and runtime environments (e.g., OpenPLC~\cite{Alves2014}). In each cycle, the controller scans all input signals, executes the control logic, and writes the analog or digital output signals. To minimize jitter, controllers utilize real-time operating systems, such as VxWorks or real-time-enabled Linux. Human operators supervise the automation using desktop applications or web-based clients.

User requirements for automation control logic typically include I/O lists, piping-and-instrumentation diagrams (P\&ID), and control narratives, written by specialized EPC (engineering, procurement, and construction) contractors with process engineering domain knowledge on behalf of the automation user (e.g., a chemical company). 
Control narratives are prose text documents (MS Word, PDF, 5-500+ pages) that specify automation requirements in natural language, referring to tag names from the I/O list and occasionally explaining structures from P\&IDs. EPCs write control narratives independently of a particular control system, allowing them to be used to solicit bids from different automation vendors.

\begin{figure}[!htbp]
\noindent\fbox{%
  \parbox{0.95\linewidth}{%
``The ratio control system is interlocked with the reactor level transmitter LT-104 and the vent scrubber VS-107 to ensure safe operation. LT-104 provides a permissive signal to FFIC-102, allowing acid flow only when the reactor level is above 20\% and below 90\% of its capacity.''
  }%
}
\caption{Control Narrative Text Snippet Example}
% \label{fig:narrative-snippet}
\end{figure}

Control engineers read such narrative texts (Fig. 3) when implementing the control logic. In the example, the control engineer needs to instantiate a function block for the level transmitter LT-104 and for the ratio controller FFIC-102, and then connect them so that a high or low alarm raised by the level transmitter stops the controller and, consequently, stops the inflow to the reactor, preventing underflow or overflow. In many cases, the specifications are more complex and require recognizing particular control strategies (e.g., ratio control, cascade control) and considering intricate inter-dependencies between the different devices. 
Control engineers are trained in vendor-specific libraries and can translate user requirements into control system specifics.% concepts (e.g., special alarm handling, special ways of processing signals).

As prose texts, control narratives may be inexact, inconsistent, and incomplete, complicating the task of the control engineer. EPCs write them in local languages and without standardized formatting or structure, which makes them convenient and cheap to create. Before the advent of Generative AI, attempts to perform information retrieval on control narratives (e.g., ~\cite{Jetley2024}) were still limited by the heterogeneity of formulations used, and thus focused only on extracting specific information items (e.g., setpoints), rather than the entire content. A few large automation customers utilize more formalized notations to specify their requirements~\cite{Drath2018,Brendelberger2013}, but even these are often accompanied by prose text control narratives.

\section{Spec2Control: Conceptual Approach}
\label{sec:spec2control}
Spec2Control's goal is to automate the previously described control logic engineering process as much as possible, thereby reducing engineering costs. We first describe Spec2Control conceptually, supported by a running example, before sketching a prototypical implementation.

\subsection{Running Example}
Consider the control narrative of a chemical production plant that manufactures ammonium nitrate, a fertilizer. Section 2 of the control narrative describes parts of the desired automation for the included neutralization reactor, referencing 10 tags related to components such as level transmitters, flow controllers, and flow valves.  

\begin{figure}[!htbp]
\center
  \includegraphics[width=\columnwidth]{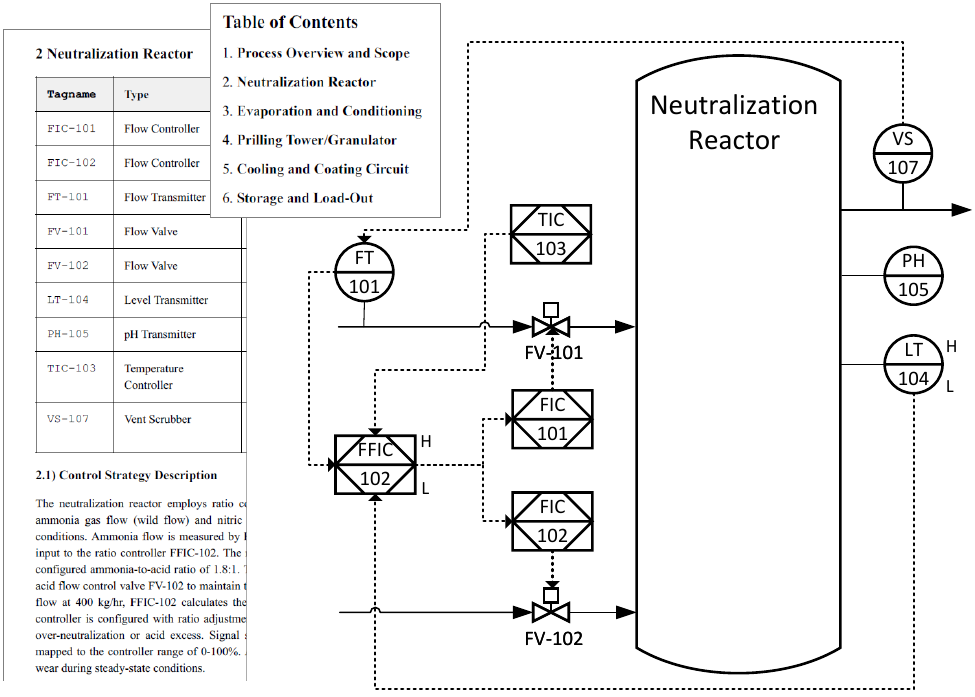}
  \caption{Running Example: Control Narrative Section on Neutralization Reactor in Ammonium Nitrates Chemical Production Plant} 
\label{fig:running-example}
\end{figure}

Figure~\ref{fig:running-example} provides a graphical P\&ID of the text for illustration purposes in the context of this paper. This graphical representation is not part of the actual specification and is not available to the LLM for FBD generation, as it exclusively relies on text. Circles indicate sensors, butterfly shapes indicate valves, and squares indicate control functionalities. The text describes a ratio control strategy involving a ratio controller (FFIC-102) with two flow controllers (FIC-101, FIC-102) regulating the inflow valves (FV-101, FV-102) of the reactor. Interlocks, permissives, and alarm signals needed in the plant segment are also described and represented in the graphics by dashed lines. A control engineer needs to map the described devices to matching function blocks from a library and then connect and parametrize these blocks. The result will be an FBD. This task typically requires 1-2 hours of manual work per section and is followed by extensive testing.

\subsection{Control Logic Generation Method}
\begin{figure*}[!htbp]
\center
  \includegraphics[width=\textwidth]{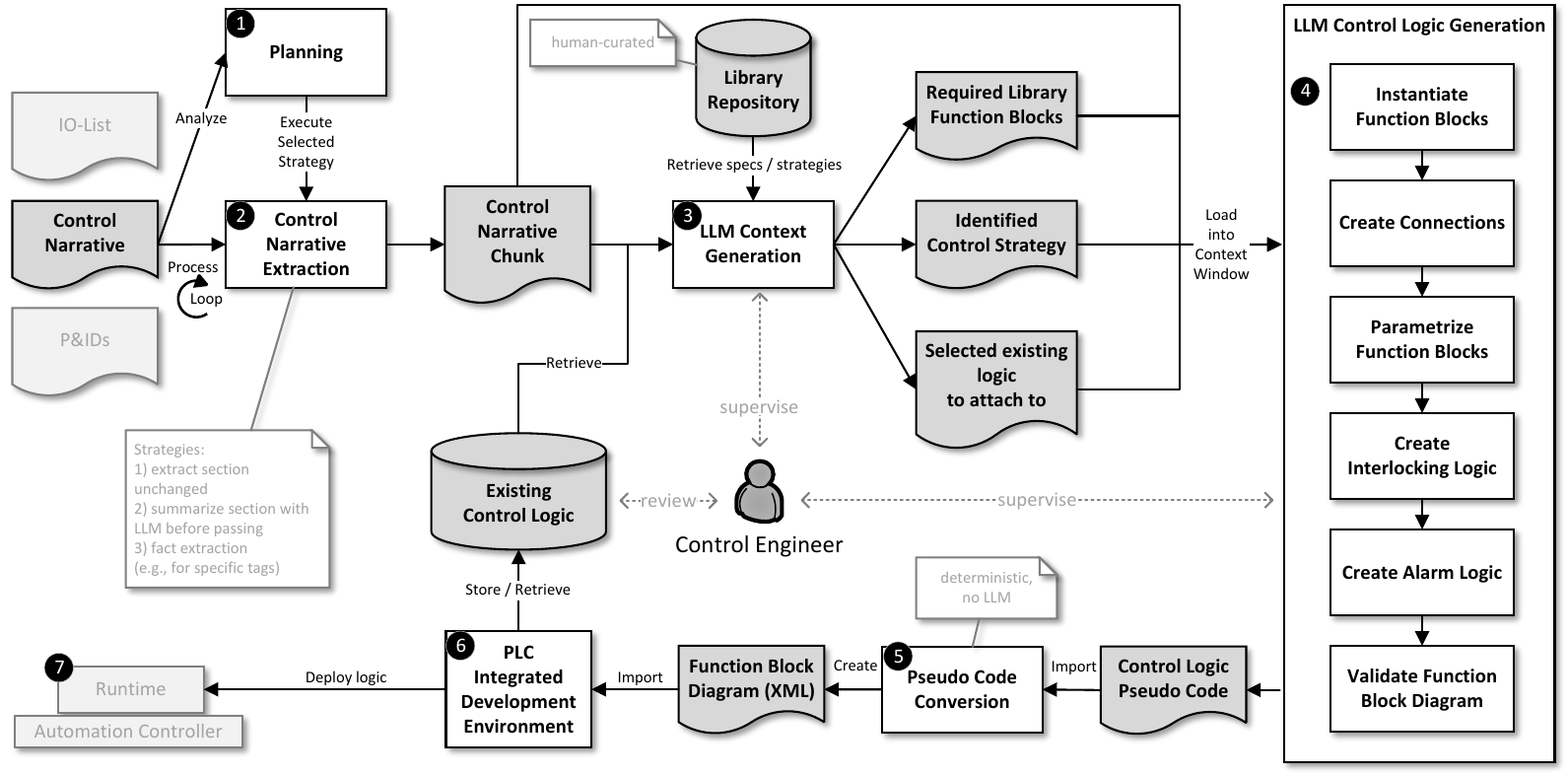}
  \caption{Spec2Control Control Logic Generation Method: control narratives serve as input, are processed in seven steps by LLM workflows and deterministic tools, resulting in control logic deployed onto an automation controller.} 
\label{fig:method}
\end{figure*}

Fig.~\ref{fig:method} illustrates the Spec2Control method, which begins with a narrative and culminates in FBD control logic deployed on an automation controller. The method consists of seven steps, where LLM agents execute the first four steps under the supervision of a control engineer. 

\textbf{Step \bcirc{1} (Planning):} First, a planning procedure analyzes the control narrative provided by the control engineer to determine a suitable chunking and processing strategy. In practice, control narrative sizes span from only 1-2 pages to more than 500 pages, in which case chunking is essential to avoid overloading LLM context windows. Chunks are ideally self-contained, semantically coherent sections mappable to individual diagrams. Often, the user-defined sections provide a good template for chunking; however, in some cases, sections are ill-defined or need to be combined for coherence. Our current prototype supports 1) processing the entire document, 2) chunking by sections, and 3) chunking by user selection. Future additions could enhance the chunking strategy selection process by utilizing LLM prompts and human revision. For the running example, chunking by sections yields semantically coherent results and is the selected strategy.

\textbf{Step \bcirc{2} (Control Narrative Extraction):} Based on the selected chunking strategy, this step extracts parts of the control narrative and feeds them one by one to the subsequent workflow. The step also encapsulates the handling of different input file formats (e.g., Word, PDF, and Markdown). The result of chunking is typically 2-3 paragraphs of text, possibly accompanied by tables and images. Optionally, this step can run pre-specified LLM summarization prompts on the chunk to strip excessive formatting instructions, redundancies, and overly verbose formulations. In practical cases, this can often reduce the chunk size by more than 50 percent, thus preserving more context window space for later instructions. For the running example, this step extracts Section 2 of the control narrative, which contains 3826 characters and 10 referenced tags. 

\textbf{Step \bcirc{3} (LLM Context Generation):} Before actually generating control logic, Spec2Control establishes an appropriate context~\cite{Mei2025}. Context generation itself is a multi-step workflow with several LLM prompts that gather 1) required function block definitions, 2) required control strategies, and 3) previously existing control logic to connect to. 

A typical control narrative chunk requires only a subset of the available library block types, and from these, only a subset of the available input and output parameters. Via LLM prompts on the current narrative chunk, Spec2Control identifies the required type definitions (e.g., PID\_BASIC), which include function block name, input/output parameters (e.g., SP for set point), their types (e.g., BOOL, REAL, etc.), and descriptions (e.g., "setpoint for the PID control loop"). Retrieval-augmented generation~\cite{Gao2023} can support this step, as shown in previous work~\cite{Koziolek2024}, but this usually requires specially prepared source documents from a library repository. For example, the section in the running example requires, among others, a function block type ANALOG\_IN with input parameters for high and low alarms to integrate the level transmitter LT-104 into the DCS (Fig.~\ref{fig:running-example}).

The step selects a suitable control strategy for the given chunk from a list of pre-specified strategies. There are a dozen popular control strategies, which already cover more than 95 percent of the typical process control cases~\cite{King2016}. These include, for example, PID, cascade, feedforward, split-range, or duty/standby control. Function block libraries support these strategies and induce specific block composition patterns, which this step loads into the context. This information helps the LLM to determine the required function block connections for the narrative chunk. For cases where a control strategy is challenging to detect or where a chunk includes overlapping strategies, Spec2Control provides confidence levels for the possible strategy options. It optionally involves the human user in making a decision. For the running example, Spec2Control autonomously selects the ratio control strategy, which here requires a RATIO\_CONTROL block and two PID\_BASIC control blocks with specific connections between the blocks, as it controls the ratio between the two reactor inflows.

Finally, step 3 also loads selected existing control logic into the context. For example, function blocks may have already been instantiated in the control logic from previous generation steps, using so-called bulk engineering tools, or through manual work. Especially for extensions of existing automation systems, this step is crucial to enhance the existing implementation seamlessly. Spec2Control searches in a repository of existing control logic based on the tag names in the current narrative chunk, and retrieves related function block instantiations and connections. For example, in the running example, the PID\_BASIC block for temperature controller TIC-103 was already available. Therefore, the Spec2Control appends its definition with connections to the context, so that subsequent steps do not need to recreate the block.

\textbf{Step \bcirc{4} (LLM Control Logic Generation):} With the required context (i.e., block types, composition patterns, existing logic) in place, the following workflow can now generate the actual control logic for the current control narrative chunk. There can be different variants of the workflow, e.g., for the different IEC 61131-3 languages (e.g., ST, FBD, SFC), for IEC 61499, or vendor-specific notations. For the scope of this work, we focus on FBDs; previous work, for example, has demonstrated the generation of ST~\cite{Koziolek2023}. 

Spec2Control divides the FBD generation into several steps, each supported by individual LLM prompts, which enables focusing the LLM processing on smaller subtasks, thereby producing more reliable outcomes. First, it instantiates block types and assigns them names based on the tag names. Then it creates block instance connections according to the spec and the identified control strategies. Parameters, such as alarm levels or setpoints, are added to the blocks before two additional steps deal with interlocking logic and alarms. Interlocks connect the states of two devices (e.g., a high alarm causes a pump to stop). In contrast, alarms define thresholds for additional logic or human operator monitoring (e.g., a reactor temperature maximum of 190 degrees Celsius). Finally, a validation step receives the output of all previous steps and checks for well-formedness and completeness using another LLM call.

\begin{figure}[!htbp]
\center
  \includegraphics[width=\columnwidth]{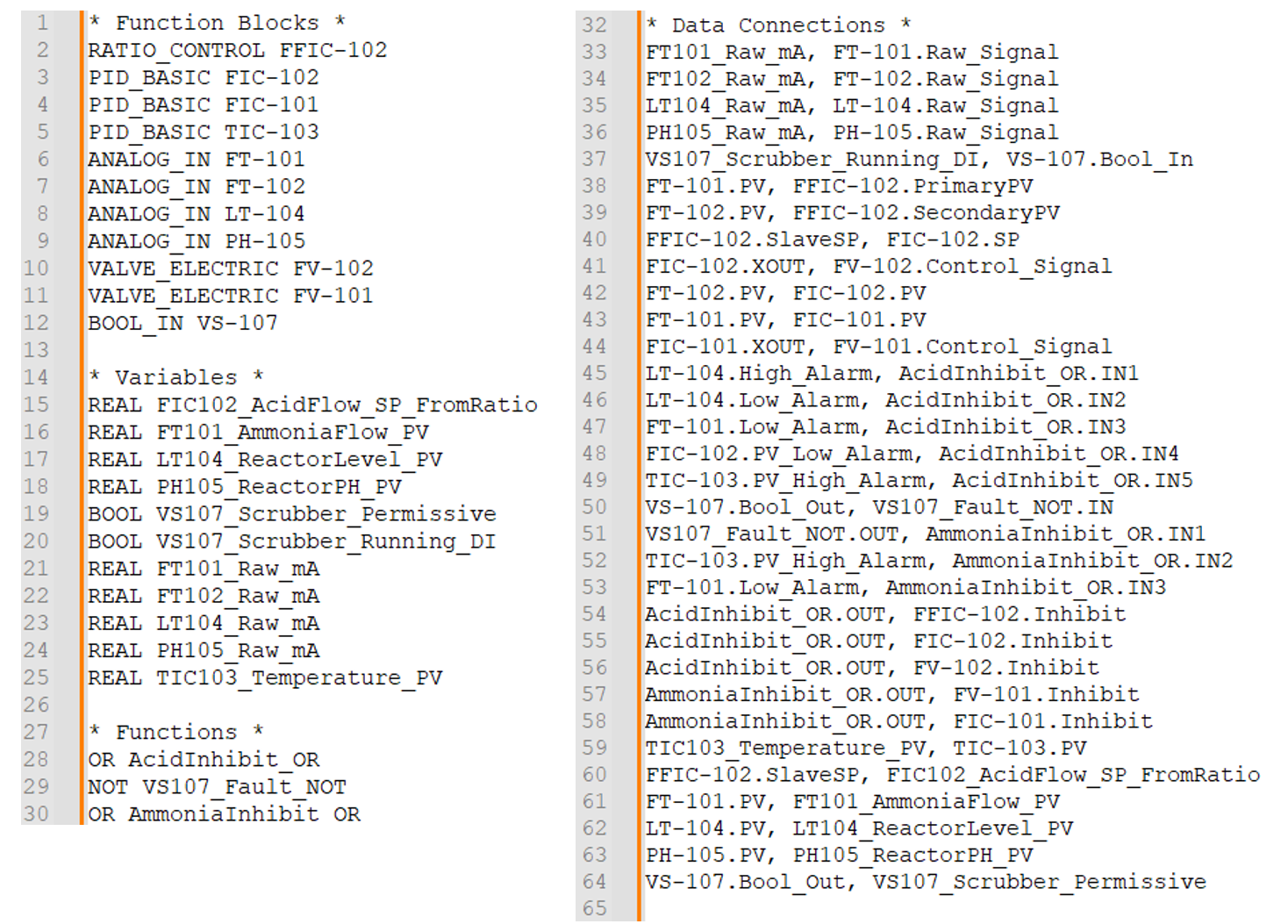}
  \caption{Spec2Control's LLM FBD output notation (``pseudo-code'') for the neutralizer reactor section of the ammonium nitrites chemical production plant (excerpt).} 
\label{fig:pseudo-code}
\end{figure}

Spec2Control defines a streamlined pseudo-code notation for FBDs, which focuses the LLM solely on outputting the essential instantiation, connection, and parametrization of function block types (running example in Fig.~\ref{fig:pseudo-code}). This notation does not include graphical coordinates, sizes of function blocks, or coordinates for edge routing, as this information is added in a subsequent step using an auto-layout algorithm. For the running example, the result of the control logic generation step is an FBD with 11 function block instances, three functions, 11 variable definitions, 32 connections, and 24 parameters.

\textbf{Step \bcirc{5} (Pseudo Code Conversion \& Layout):} Spec2Control parses the pseudo-code notation, converts it into a data format required for the PLC IDE, and generates graphical coordinates for the FBD blocks and connections. This step does not utilize an LLM but instead employs deterministic code. There are vendor-specific proprietary formats and standardized formats (PLCOpen XML). This step enables us to adapt the method to different PLC IDEs by simply replacing the conversion logic, while leaving steps 1-4 unchanged. A prerequisite for such an adaptation to work is that the library block types are available in the target PLC IDE. Auto-layout for FBDs requires a custom hierarchical graph layout with horizontal orientation and backward looping~\cite{Tamassia2013}. For the running example, this step yields an FBD in OpenPLC's implementation of the PLCopen XML schema. 

\textbf{Step \bcirc{6} (PLC Integrated Development Environment):}
When loaded into a PLC IDE, the FBD can be rendered graphically and displayed to the user. Fig.~\ref{fig:fbd} in Section 4.2 shows an excerpt of the FBD auto-layouted and rendered for the running example FBD in OpenPLC. If required, the user can now use the IDE to correct the FBD, to add more logic beyond the narrative chunk, and to test the implementation. Additional control logic, for example, can come from P\&IDs or other sources, which is outside the scope of this work. LLM-generated test cases can support the validation~\cite{Koziolek2024}.

\textbf{Step \bcirc{7} (Runtime):}
Once complete and tested, the user deploys the control logic to an automation controller, which is network-connected to the sensors and actuators, and cyclically executes the logic.

A control engineer supervises the entire control narrative generation process and is looped in for additional input in case the executed workflows encounter low-confidence situations or ambiguities. Spec2Control can integrate with existing bulk engineering tools~\cite{ABB2019}, which in some cases can already pre-populate PLC IDEs with unconnected function block instances based on I/O lists. 

\subsection{Prototype Implementation}
Different LLM application frameworks can support the implementation of Spec2Control. We chose to build a prototype based on LangChain~\cite{Annam2025}. The prototype is connected to ABB control engineering tools~\cite{ABB2019} and loaded with specifications of several proprietary ABB function block libraries. For independent reproduction and evaluation of the concepts, we have now also implemented an adapter for the open-source PLC IDE OpenPLC~\cite{Alves2014}. This adapter enables interested users and researchers to evaluate the approach without exposing proprietary ABB information.

The prototype~\cite{Koziolek2025} follows the steps described previously. For Step 1 (Planning), the user manually selects a narrative chunking strategy; a more dynamic strategy selection is under development. In Step 2 (Control Narrative Extraction), Spec2Control utilizes PyPDF to split PDF files and extract the raw text. The user can optionally integrate prompts for summarizing control narrative chunks to reduce the token count. For Step 3 (LLM Context Generation), we utilize LangChain to instantiate prompt templates and integrate LLM calls. We use OpenAI LLM models hosted on Azure AI Foundry (currently GPT-5). The user can configure the specific model, as Spec2Control is LLM-model agnostic and thus can benefit from the best available models. The merging of existing control logic is currently under development. 

In the current prototype, the LLM Control Logic Generation (Step 4) for FBDs is a linear prompt chain, which is easy to test. There are other prompt workflows for generating SFC and ST. We crafted the included prompts to support typical IEC 61131-3 logic. The prompt chains result in pseudo-code, which Spec2Control converts into ABB proprietary formats or PLCopen XML (Step 5). As OpenPLC does not support auto-layout of FBDs, we have implemented this functionality in our prototype. A FastAPI exposes all workflow steps. 

\begin{figure}[!htbp]
\center
  \includegraphics[width=\columnwidth]{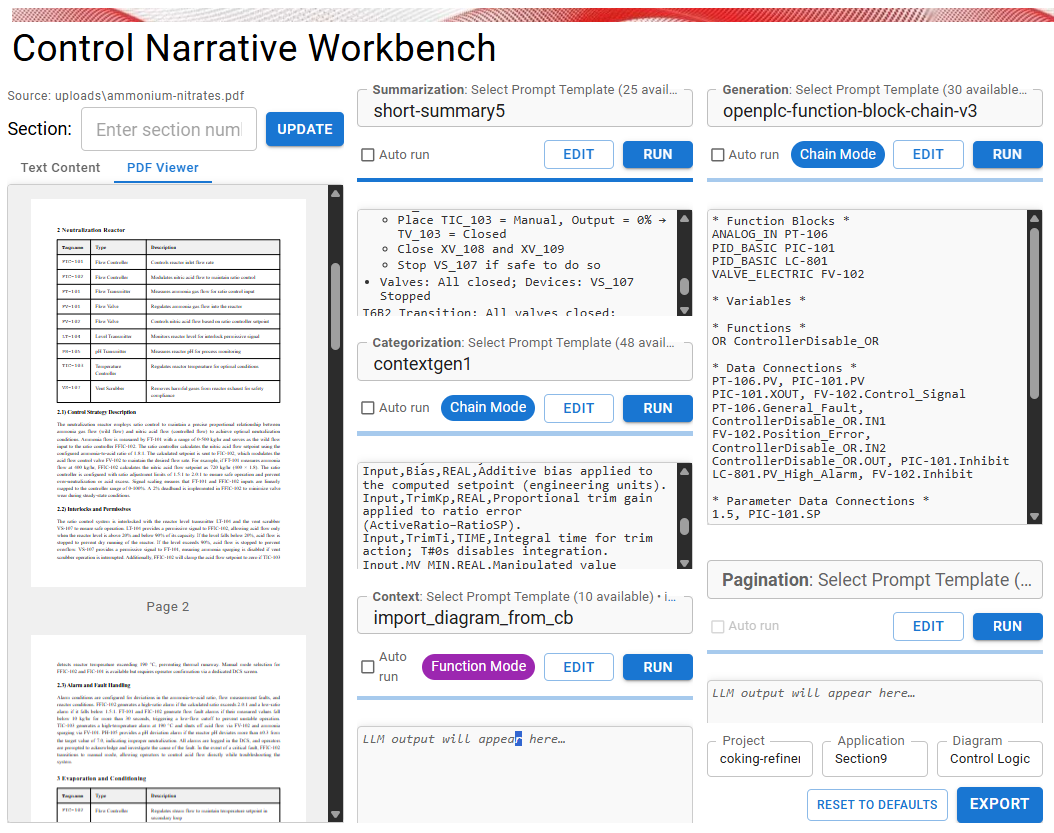}
  \caption{Control Narrative Workbench: an interactive front-end for Spec2Control to assist in prompt engineering and LLM supervision} 
\label{fig:cnw-screenshot}
\end{figure}

For the frontend, we have created a purpose-built user interface called ``Control Narrative Workbench'' using ReactJS, where control engineers can load control narratives, perform prompt engineering, configure workflows, and monitor LLM outputs (Fig.~\ref{fig:cnw-screenshot}). A command-line utility allows users to execute the steps unattended and process entire control narrative documents in batch mode.

\section{Case Studies}
A GQM (Goal-Question-Metric) approach~\cite{Caldiera1994} structures the evaluation of Spec2Control. The goal was to determine the feasibility of automated control logic generation from natural language specifications. Derived from the goal, we asked three questions:
\begin{enumerate}
\item \textbf{Question Q1:} How many human interventions are required, and how complex are they? Derived metric Q1M1 counts the number of human interventions to produce FBD control logic for a given control narrative chunk. As metric Q1M2, we estimate the rework time required for generated FBDs resulting from erroneous LLM-generated content. As metric Q1M3, we estimate the time needed for review and correction to achieve confident semantic correctness.
\item \textbf{Question Q2:} What is the quality of the LLM-generated control logic? Notice that there are different ways to implement the functionality defined in a control narrative chunk. For example, other naming conventions are valid, and the LLM may combine complex Boolean expressions in various valid ways. To address this variability, we created a baseline of FBDs, reviewed by domain experts, to serve as a reference for evaluation. Since exact structural matches are not a reliable indicator of correctness, we instead assessed each LLM-generated output using the following metrics compared to the baseline:
\begin{itemize}
\item Q2M1: percentage of correctly identified control strategies
\item Q2M2: percentage of correctly instantiated function blocks relevant to the control strategy

\item Q2M3: percentage of correct control strategy connections 
\item Q2M4: percentage of correctly configured alarm parameters
\end{itemize}
\item \textbf{Question Q3:} What is the cost/benefit ratio of an automated vs. manual control engineering approach? Metric Q3M1 is the percentage of LLM construction time plus LLM-induced rework time divided by the estimated manual construction time. Metric Q3M2 quantifies the average LLM costs to process a control narrative chunk in USD.
\end{enumerate}
% make estimation based on the complexity of the fbd, with number of data connections superlinearly increasing the efforts due to the sophistication level
Measuring the metrics on representative reference test cases helps assess the feasibility and usefulness of the approach, providing ABB control engineering units with data for initiating AI adoption across worldwide engineering centers.

\subsection{Test Setup}
To collect data for the metrics, we analyzed dozens of \textbf{control narratives} from ABB automation customers in recent and past ABB automation projects. These control narratives ranged in size from 5 to 500+ pages or more and concerned various applications, including offshore wind farms, wastewater treatment facilities, and power plants. Such control narratives are proprietary, about intellectual properties, and thus, we cannot publish them. We therefore created an LLM-based control narrative generator that can synthesize realistic control narratives utilizing process descriptions from textbooks or based on P\&ID images. 

Using the generator, we created 10 non-proprietary control narratives across various industrial automation use cases to capture a wide range of control strategies and required equipment. A large-scale industrial plant inspires each control narrative. Three of the narratives originate from P\&ID images, while seven narratives come from textbook seed texts. We ensured a detail level comparable to commercial control narratives, which were cross-checked by experienced control engineers. The control narratives comprise up to 10 sections, each spanning 1-3 pages, and collectively covering more than 50 tags (sensors, controllers, actuators). Each section contains a tag name table and describes the main control philosophy for a plant segment (e.g., ``vacuum distillation unit'') in prose text referring to the tag names. Each section also provides subsections for interlocks and alarms. The non-proprietary control narratives enable a transparent test setup outside corporate borders, which automation experts and researchers can validate.

\begin{table}[!htbp]
\center
  \includegraphics[width=0.9\columnwidth]{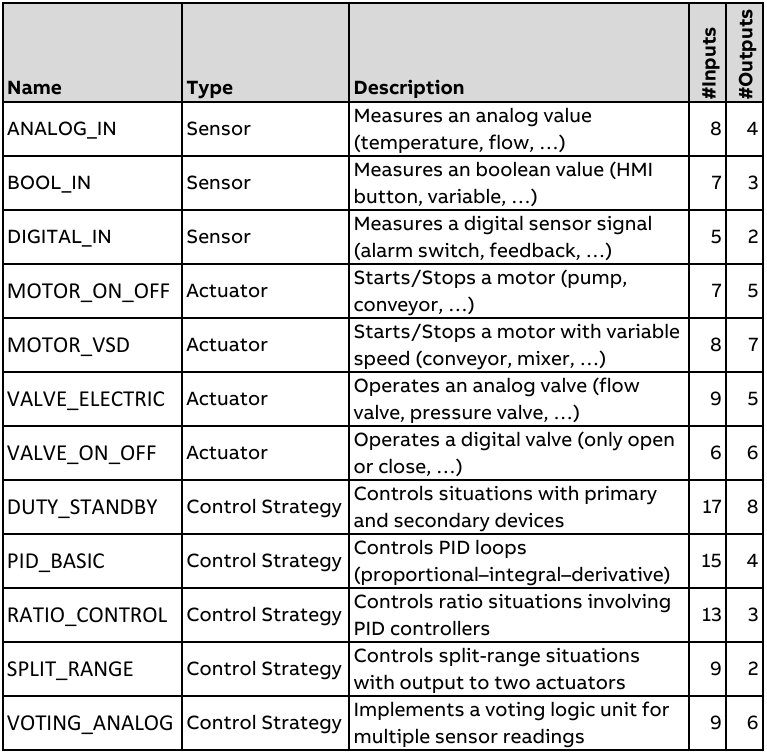}
  \caption{Function Blocks in BASIC\_LIB library} 
\label{tab:basic-lib}
\end{table}

We also analyzed more than a dozen ABB control \textbf{function block libraries}, which are used in different industry domains (e.g., metals, chemicals, power production, etc.) and in other regions (e.g., Europe, US, Asia). Often, these libraries contain 15-20 major function blocks with 50+ input and output parameters, in addition to smaller supporting function blocks. Function block libraries are also proprietary and carry company-internal IP. Therefore, we defined a new library (``BASIC\_LIB'') inspired by the commercial libraries. It contains 13 function blocks, which, for example, convert digital and analog input signals, implement control algorithms, and steer valves and pump motors (Table~\ref{tab:basic-lib}). Each block defines up to 10 input and 10 output parameters, along with their type and textual description. While some commercial function blocks can have more than 50 parameters, in many situations, control engineers only use a few of these parameters, while the others encapsulate specialized functionality. Therefore, the function blocks in the BASIC\_LIB capture the most typical parameters, which cover 80-90 percent of the usual functionality. For validation purposes, we defined the blocks as OpenPLC user-defined blocks, allowing OpenPLC to instantiate them. Besides the BASIC\_LIB function blocks, our LLM control logic generation workflow can also instantiate standard IEC 61131-3 function blocks, including, for example, boolean expressions, arithmetic calculations, or timers.

We created a \textbf{baseline} implementation of the FBDs for the 65 sections of our 10 control narratives. Subject matter experts manually reviewed these FBDs and verified their correct functionality and completeness. They found that the BASIC\_LIB blocks could not explicitly address all specific details of the control narratives, such as handling particular operator interactions via UI elements. Furthermore, they found that the BASIC\_LIB interlock handling with inhibit pins was slightly simplified compared to some commercial libraries, but not unrealistic. The subject matter experts, however, did not consider these gaps as critical and focused on reviewing the main control strategy implementations and alarm logic. We then used this baseline implementation to compute the metrics Q2M1-Q2M4 from test runs and to make the effort estimations for Q3M1.

The \textbf{test execution} comprised running the workflows for extraction, context generation, control logic generation, and pseudo-code conversion. We used our CLI tool to run the tests in a batch style against the entire control narrative dataset. OpenAI GPT-5 with auto-reasoning needed up to 1 minute for individual prompts and approximately 1 hour for a complete control narrative. After test execution, scripts computed the metrics Q2M1-Q2M4, whose results we manually reviewed.

\subsection{Results with Open Dataset}
\begin{figure*}[!htbp]
\center
  \includegraphics[width=\textwidth]{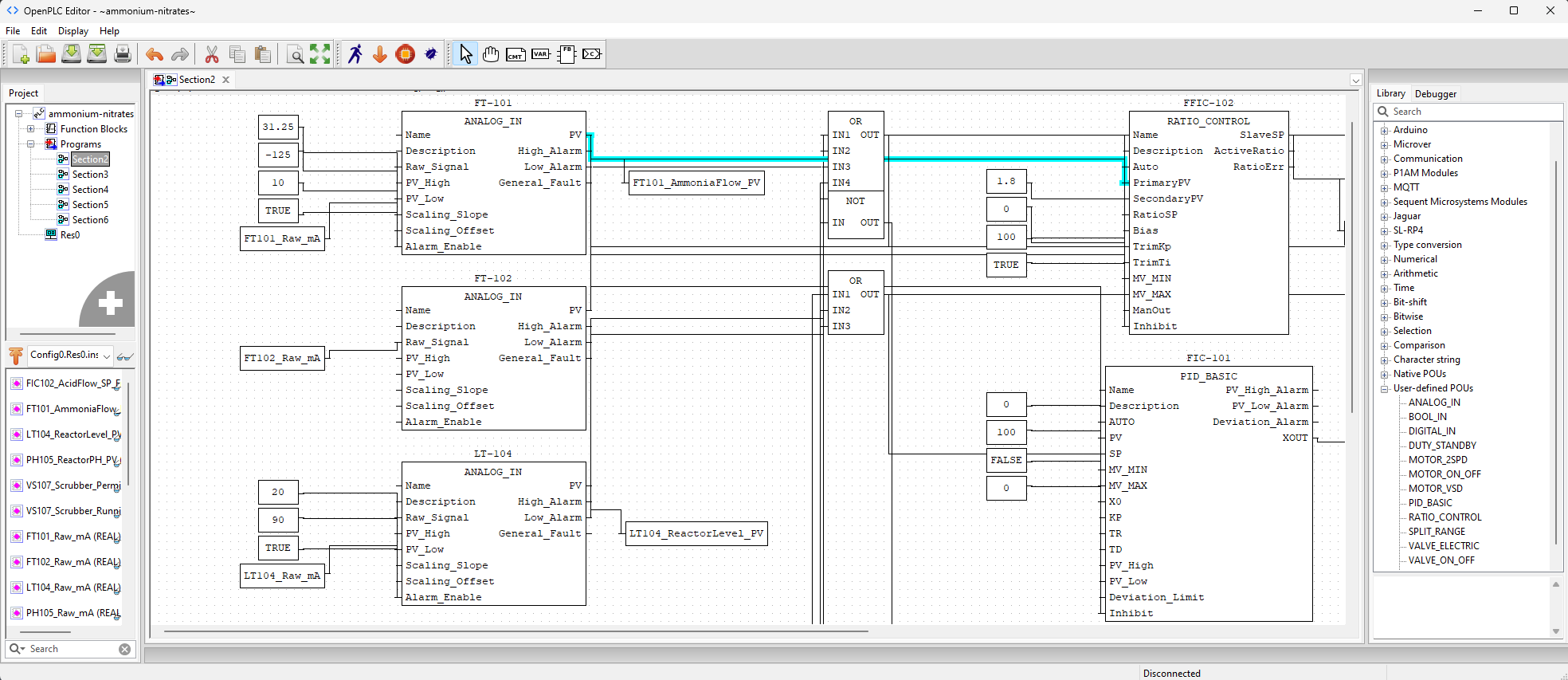}
  \caption{Excerpt of an LLM-generated Function Block Diagram (FBD) for the Ammonium Nitrates Plant (Section Neutralization Reactor) in OpenPLC. The LLM has instantiated the required function block types from a user-defined library and connected and parametrized them according to the detected control strategy.}
\label{fig:fbd}
\end{figure*}

For illustrative purposes, Fig.~\ref{fig:fbd} shows an excerpt of an LLM-generated FBD for the running example described in Section~\ref{sec:spec2control}. Due to space constraints, we cannot display the entire FBD in this paper; however, the full 65 FBDs for the complete dataset are available in the paper's supplementary material~\cite{Koziolek2025}. In the top left corner of the FBD, the figure shows an ANALOG\_IN function block named ``FT-101'', which the LLM has instantiated for the flow transmitter FT-101. This function block has an output ``PV'' (for process value), which drives the RATIO\_CONTROL block FFIC-102, according to the ratio control strategy detected by the LLM in the control narrative section (highlighted connection in light blue). Several functions ``OR'' and ``NOT'' implement interlocking control logic involving Boolean expressions. The LLM correctly configured the alarm levels of the ANALOG\_IN block ``LT-104'' with values of 20 and 90 for the low and high alarm thresholds of the signal. The FBD control logic will thus ensure that the reactor inflow valves are closed when the tank level rises above 90 percent. The complete diagram includes 11 function blocks, three functions, 11 variables, 32 data connections, and 24 parameters.

Our unattended test runs required no human intervention (\textbf{Q1M1} = 0), which demonstrates that the LLM workflows and prompts are defined with a sufficient level of detail and generality to handle the various control narrative specifications. Context generation was fully automated, as well as diagram creation and auto-layout. However, our tests did not yet involve merging prior control logic or utilizing rare parameters of function blocks, which may require human input in some instances. Furthermore, the control narrative texts were reviewed before FBD generation. In contrast, other control narratives may contain inconsistencies that require human intervention, including clarifications with the narrative's author. An LLM could flag such inconsistencies before attempting FBD generation.

The generated FBDs are complex and require careful inspection at the moment. A PLC IDE can often already indicate obvious errors (e.g., type mismatch in connections, or disallowed characters) to the user. Mismatches in function block type names could prevent the IDE from instantiating them; however, in our tests, such mismatches did not occur. Therefore, we foresee the time for checking syntactical errors of the LLM-generated FBDs as almost negligible per FBD ($\textbf{Q1M2} \approx 5$ min).

Checking semantic correctness is challenging, especially with complex ``glue logic'' such as Boolean expressions, timers, and arithmetic. Human-crafted FBDs require functional checks through peer reviews by control engineers. Based on tests, we estimated review times for FBDs by complexity, allocating an average time per element (e.g., 1 minute per connection). Review times across the dataset ranged from 19.4 to 114.7 minutes per section, with a median of 42.8 minutes (\textbf{Q1M3} = 42.8 min).

\begin{figure}[!htbp]
\center
  \includegraphics[width=\columnwidth]{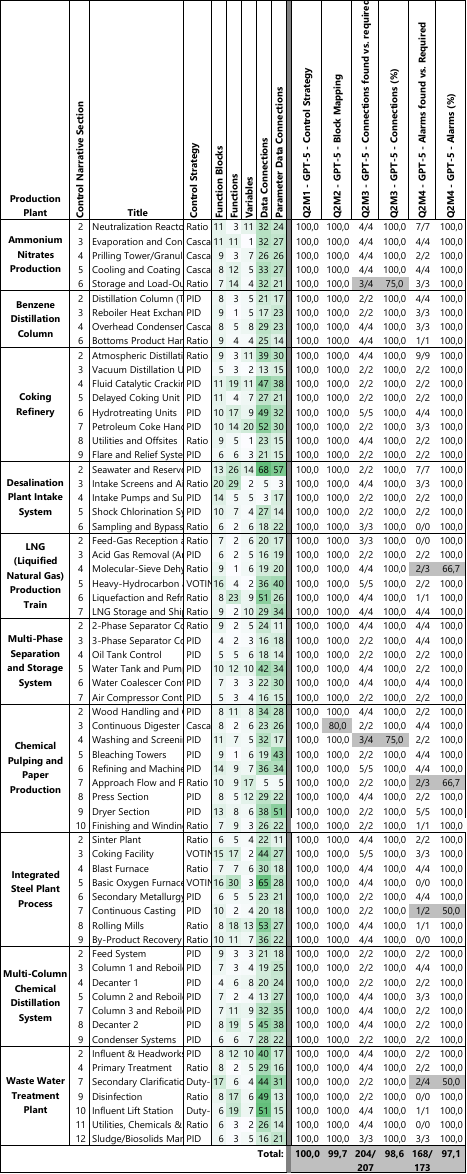}
  \caption{Open Dataset with 10 control narratives and 65 sections: statistics on the baseline FBDs (left-hand side) and metrics Q2M1-Q2M4 based on test runs (right-hand side).} 
\label{fig:results}
\end{figure}

Fig.~\ref{fig:results} provides an overview of all the control narrative sections in our dataset. It illustrates the diverse range of production plants (e.g., chemical, water, paper, steel) and the realistic segments within these plants, presented in linear sequence across the different sections. There is a mix of various control strategies, with ``PID control'' being the most prevalent strategy, which aligns with experience from practice. The statistics for the baseline FBDs indicate that the control narrative sections required 4-20 function block instantiations and 3-68 connections per FBD. 
%- some sections left out, because the narrative text contained errors and inconsistencies

Fig.~\ref{fig:results} depicts the measurements for Q2M1-Q2M4 to assess the quality of the LLM output compared to the manually reviewed baseline for each tested control narrative section. As there are different valid ways to implement the narrative specifications in an FBD, the LLM test outputs will never be identical to the baseline. Our metrics assess whether the LLM-generated FBDs accurately recognize and implement the required control strategies, and whether the LLM correctly maps important alarm parameters from the specification to the function block. GPT-5 consistently identified the control strategy needed in all cases flawlessly (\textbf{Q2M1}= 100\%). 

Q2M2 assessed whether transmitters and actuators from the control narrative were correctly mapped to BASIC\_LIB function blocks, in line with the identified control strategy. Our automated scripts indicated several mismatches between the block mappings of the baseline and the test runs. However, upon reviewing these mismatches, we found that most of them stemmed from either incompleteness or ambiguities in the control narrative, which allowed for different valid ways of implementing the FBD. A detailed explanation of each mismatch is in the supplementary material~\cite{Koziolek2025}. Only for the continuous digester of the Chemical Pulping and Paper Production (section 3), the LLM-generated FBD missed mapping the control valve LV-103 to a VALVE\_ELECTRIC function block. For the other sections, GPT-5 instantiated the control strategy-related function blocks correctly (\textbf{Q2M2}=99.6\%).

Q2M3 measured the correct creation of connections related to the identified control strategy. We omitted comparing additional ``glue logic'' connections to the baseline, as there are many degrees of freedom in the implementation that would result in an exact match only by chance. Except in two cases, GPT-5 generated the control strategy connections in the FBD correctly (\textbf{Q2M3}=98,6\%). In one case, the LLM generated an erroneous connection to a ratio controller. In the other case, the LLM routed a feedback signal from a valve back to a PID controller, which our subject matter experts considered incorrect. 

Q2M4 measured the correct mapping of important alarm parameters from the specification to the FBD. In 5 of 173 cases, the LLM-generated mapping to the FBD resulted in a mismatch with the baseline FBD, as the LLM created values that were not present in the control narrative or missed specified values. However, in 168 out of 173 cases, GPT-5 mapped the alarm parameters correctly (\textbf{Q2M4}= 97.1\%).

Qualitative findings support the quantitative metrics: The LLM created necessary function block instantiations that were not explicitly mentioned in the control narrative, indicating that the narrative was slightly incomplete. Its additions matched what a control engineer would do. The LLM also inferred parameters such as scaling slopes and alarm levels correctly. Although the narrative specified interlocking logic in separate sentences, the LLM formulated them into correct Boolean expressions. Our limited BASIC\_LIB library complicated the implementation of some safety features (e.g., permissive ranges), leading the LLM to devise custom logic or omit these features. Enhancements in BASIC\_LIB are needed.

For quantifying the cost/benefit ratio of Spec2Control, Q3M1 divides the LLM construction time (Q1M1) plus LLM-induced rework time (Q1M2) by the manual construction time of an FBD. This ratio excludes the time for checking semantical correctness (Q1M3), which is still required both with and without Spec2Control. Manual construction time is estimated as a multiple of Q1M3, as the manual creation of the FBD requires understanding and interpretation of the control narrative, as well as manual work to use the PLC IDE, depending on the complexity of the FBD. With a factor of 2 - 3 on Q1M3 (median 42.8 minutes), the manual creation of an FBD for a single control narrative section would require between 85.6 and 128.4 minutes. With LLM-construction time assumed zero (no human labor due to Q1M1=0) and Q1M2 estimated as 5 minutes human labor, this would lead to a time saving between 94.2\% and 96.1\% (\textbf{Q3M1}$\approx 95\%$).

While the FBD generation leads to time and cost savings due to shorter human labor, the generation induces monetary costs for LLM token usage. From our test runs, Azure AI Foundry monitoring tools showed that generating an FBD for a single control narrative section required approximately 25,000 input tokens and 60,000 completion tokens. According to current pricing, Azure GPT-5 token costs are 1.25 USD per 1 million input tokens and 10 USD per 1 million completion tokens. Therefore, the costs for processing a control narrative chunk and generating an FBD are 0.031 + 0.6 = 0.631 USD (\textbf{Q3M2}$\approx$ 0.63 USD). Compared to human labor costs, the LLM costs are thus negligible. %Costs may, of course, vary in other contexts, such as when local LLM hosting is required (e.g., due to privacy concerns).

\subsection{Results with Commercial Data}
We also tested Spec2Control in several commercial ABB automation projects, including a desalination plant with multiple pumping stations, a water supply system for an offshore converter station, and an asphalt production facility. These tests used different ABB libraries and generated ABB's proprietary version of FBDs. We found that experiences with using libraries and best practices for composing the pre-specified blocks require interviews with experienced control engineers. Context generation may differ significantly between different libraries, meaning that each library requires a distinct prompt workflow.

These tests used various strategies to segment control narratives. Some short narratives were processed in one shot, while others required manual selection of coherent segments. For standalone chunks focusing on a single strategy without sequential logic, we achieved the same LLM output quality as the open dataset. Control engineers are keen on the results due to the tool's integration with their systems and the familiar notation it generates. A software team is refining the Spec2Control prototype into a commercial product.

\section{Threats to Validity}
After analyzing our results, we discuss potential threats to their validity. The \textbf{internal validity} refers to causal correctness, i.e., whether the outcome (generated FBDs) was really caused by the treatment (using Spec2Control/LLM) or whether other factors interfered with the results. In our case, LLM non-determinism could affect the generated FBDs. We noticed, for example, slight variations in the naming of variables or in the composition of glue logic across different GPT-5 runs. Still, the non-determinism rarely led to syntactical errors. The metrics Q2M1-Q2M4 for the open dataset remained stable across runs, indicating that our LLM workflows sufficiently control LLM token generation. 

Another threat to internal validity is the missing FBD verification, e.g., based on a simulation of sensors and actuators. The safety of Spec2Control-generated FBDs relies on three principles: First, Spec2Control uses pre-validated function blocks, instantiated from extensively tested libraries. Second, unlike general code generation, FBD generation is limited to instantiating predefined block types, creating connections, setting numerical parameters, and composing glue logic, which reduces potential failure modes. Third, industrial practice already requires a layered verification procedure, where logic is simulation-tested before deployment, undergoes factory acceptance tests (FAT)~\cite{Hollender2010}, and also relies on runtime monitoring. Thus, Spec2Control fits well within this existing safety framework. Our subject matter experts deemed the baseline FBDs sufficiently valid based on inspections.

The \textbf{construct validity} refers to whether the operational measures (e.g., test cases, function blocks, LLM) truly represent the constructs (e.g., customer narratives, commercial libs, engineering tools) that shall be studied. While we generated the control narratives for testing Spec2Control, they are based on screening many customer-written narratives and were reviewed by experienced ABB control engineers. These control narratives may not capture all the intricacies observed in practice, such as incompleteness, inconsistencies, local languages, or distorted images, which require addressing through pre-processing steps. Spec2Control currently supports the most popular control strategies (e.g., PID, cascade), but has not been tested for other, less common strategies or custom logic. The BASIC\_LIB used for testing is synthetic due to IP concerns and less complex than commercial libraries~\cite{ABB2019,Siemens2025,Beckhoff2025}. It may thus be unable to deal with any control narrative. We focused on FBD as a popular IEC 61131-3 notation~\cite{Tiegelkamp2010} and excluded sequential procedures, which control engineers usually realize with SFCs. Our tests ran with GPT-5 as a typical state-of-the-art LLM. Subject matter experts reviewed our constructs and deemed them realistic, as they accurately reflect many practical cases.

The \textbf{external validity} questions if our findings apply beyond the tested dataset. We have conducted successful spot tests of Spec2Control with customer narratives, ABB libraries, and PLC IDEs. Our LLM workflows can extend to other notations, such as SFC and ST, or generate C code. Spec2Control can adapt to various vendor-specific libraries and notations. Our tests span multiple industries and encompass primary control strategies, interlocks, and alarms. HMI functionality is not yet supported. Implementing production sequences, such as startup/shutdown, in FBDs is complex and requires Spec2Control's extension to SFCs with embedded ST.

We support the \textbf{reliability} of our study by publishing the test data and Spec2Control implementation in supplementary data for independent validation and reproduction. 

\section{Related Work}
Modern control logic engineering in industrial automation often involves notations based on IEC 61131-3, IEC 61499, or vendor-specific variants~\cite{Hollender2010,Tiegelkamp2010}. Briefly after the popularization of LLMs, researchers demonstrated the capabilities of GPT-4 to synthesize syntactically correct IEC 61131-3 ST programs~\cite{Koziolek2023,Tran2024,Xia2025}. Fakih et al.~\cite{Fakih2024} from Siemens proposed LLM4PLC. This approach utilizes fine-tuned LLMs for a Siemens variant of ST (SCL), which generates simple control logic and performs model checking in a lab environment. Liu et al.~\cite{Liu2024} introduced a multi-agent framework for ST-code generation, called Agents4PLC, which incorporates chain-of-thought reasoning and iterative refinement. Yang et al.\cite{Yang2024} developed AutoPLC, a framework for LLM-supported  ST-code generation using a database of implementation examples. Haag et al.~\cite{Haag2025} fine-tuned LLMs for the generation of ST-code, achieving a higher compilation success rate. All of these approaches used small test programs and generated output in the simple ST notation.

Researchers then continued to refine LLM-based control logic generation. Koziolek et al.~\cite{Koziolek2024} showed how proprietary information can be integrated into ST-code using retrieval-augmented generation. Another approach demonstrated LLM-based ST generation from P\&ID images, utilizing the capabilities of a multimodal LLM~\cite{Koziolek2024a}. Zhang and de Sousa~\cite{Zhang2024,Zhang2025} explored generating graphical IEC 61131-3 notations (LD, SFC) via ASCII art. Test case generation for IEC 61131-3 ST function blocks was investigated with moderate success~\cite{Koziolek2024b}. Salari et al. converted the ST-code to Python to enable the reuse of LLM-based test case generation approaches~\cite{Salari2025}. Ren et al. proposed ``MetaIndux-PLC'', a fine-tuned foundation model for control code generation, which enables the creation of more accurate and structured PLC code. Siemens, Beckhoff, and B\&R Automation created industrial copilots, mainly for interactive PLC programming in ST. So far, none of these approaches has generated IEC 61131-3 FBDs or utilized industrial-scale test samples. Therefore, a direct comparison of Spec2Control to these approaches is not appropriate since they address different use cases.

Before the advent of generative AI, researchers and practitioners had been pursuing the automation of the laborious and cost-intensive control engineering process for several decades~\cite{Koziolek2020}. One line of work tried synthesizing ST control logic directly from \textbf{object-oriented P\&IDs}~\cite{Pang2014,Arroyo2016,Koltun2018,Drath2018,Kim2022,Schoch2024}. Drath and Fay proposed the CAEX translator~\cite{Drath2006}, which applied rules on P\&IDs to detect topological patterns that could be translated into interlocking logic expressed as ST-code. Steinegger and Zoitl~\cite{Steinegger2012} explored translating industrial requirements into a reference ontology from which they generated IEC 61131-3 ST. The AUKOTON method~\cite{Haestbacka2011} mapped XML-based P\&IDs into a domain-specific model in Eclipse Ecore and then generated IEC 61131-3 programs. Gruener et al. ~\cite{Gruener2014} converted a P\&ID-based plant structure into a Neo4J graph database and used Cypher queries to generate control logic. Koziolek et al.~\cite{Koziolek2020a} showed the successful derivation of interlocking logic from P\&IDs in four large-scale case studies.

Another line of work explored \textbf{model-driven techniques} from software engineering to generate control logic. Researchers modeled manufacturing equipment using extensions to UML~\cite{VogelHeuser2005,Hussain2006,Panjaitan2006,Witsch2009,Witsch2011,Thramboulidis2011} and SysML~\cite{VogelHeuser2014} and then generated IEC 61131-3. Others proposed novel domain-specific modeling notations~\cite{Estevez2007,Lukman2013,Schumacher2013,Schumacher2013a,Schumacher2014,Alvarez2016,Julius2017} to ease the generation of IEC 61131-3 control logic. As a result of this research, one commercial PLC IDE (CODESYS) now enables UML modeling; however, many control engineers still prefer to work directly with the IEC 61131-3 notations.

\section{Conclusions}
We introduced Spec2Control, an LLM workflow that generates control logic autonomously from control narrative specifications. Spec2Control chunks control narratives, builds up context corresponding to a control library of pre-specified function blocks, and then creates function block diagrams and passes them to typical PLC IDEs for rendering. Tests with a comprehensive open dataset demonstrated that Spec2Control can, in almost all cases, generate control logic of high quality (with more than 97\% correctness compared to a baseline). The approach can save around 95\% of effort compared to the manual control logic engineering.

Programmers and control engineers in industrial automation can test Spec2Control on their own data or utilize the open dataset to refine their own LLM-based generation approaches. The open dataset is independent of Spec2Control and published in the supplementary material. Researchers can independently validate the results and further refine the approach. We are currently working with ABB control engineers to productize the tooling and further strengthen it, so that control engineers can use it in many future automation projects.

Planned near-term enhancements of Spec2Control include addressing sequential logic and translating it into sequential function charts (SFCs) with embedded structured text. Merging newly generated logic with existing logic is under development. Furthermore, the planning step for chunking the control narratives shall be automated and refined further. We will enhance the LLM workflow to make it more robust in handling inconsistencies or incomplete inputs by incorporating human user feedback. Additional user requirements, such as I/O lists and graphical P\&IDs, shall be integrated into the processing chain to make it more flexible and comprehensive. Future work with a long-term focus includes extending the approach from control logic to the generation of graphical human operator interfaces and supporting the conversion of legacy control logic into modern notations for migration projects.

% \begin{acks}
% We thank Alejandro Carrasco, Darren Schulz, Khaled Mabrouk, and Thoralf Schulz from ABB for their support of Spec2Control.
% \end{acks}

\bibliographystyle{ACM-Reference-Format}
\bibliography{icse2026}

\end{document}